\documentclass[preprint2]{aastex}

\shorttitle{Trojan Stars in the Galactic Center}
\shortauthors{Fujii et al.}

\begin{document}

\title{Trojan Stars in the Galactic Center}

%% Use \author, \affil, and the \and command to format
%% author and affiliation information.
%% Note that \email has replaced the old \authoremail command
%% from AASTeX v4.0. You can use \email to mark an email address
%% anywhere in the paper, not just in the front matter.
%% As in the title, use \\ to force line breaks.

\author{M. Fujii\altaffilmark{1,3}, M. Iwasawa\altaffilmark{2,3},
Y. Funato\altaffilmark{2}, and J. Makino\altaffilmark{3}}

\altaffiltext{1}{Department of Astronomy, Graduate School of Science, The
University of Tokyo, 7-3-1 Hongo, Bunkyo, Tokyo 113-0033;
fujii@cfca.jp}

\altaffiltext{2}{Department of General System Studies, College of Arts and
Sciences, The University of Tokyo, 3-8-1 Komaba, Meguro, Tokyo 153-8902; 
iwasawa@cfca.ac.jp, funato@artcompsci.org}

\altaffiltext{3}{Division of Theoretical Astronomy, National
Astronomical Observatory of Japan, 2-21-1 Osawa, Mitaka, Tokyo,
181-8588; makino@cfca.jp}

\begin{abstract}
We performed, for the first time, the simulation of spiral-in of a star
cluster formed close to the Galactic center (GC) using a 
fully self-consistent $N$-body model. In our model, the central
super-massive black hole (SMBH) is surrounded by stars and the star
cluster. Not only are the orbits of stars and the cluster stars
integrated self-consistently, but the stellar evolution,
collisions and merging of the cluster stars are also included. We found that an
intermediate-mass black hole (IMBH) is formed in the star cluster and
stars escaped from the cluster are captured into a 1:1 mean motion
resonance with the IMBH.  These ``Trojan'' stars are brought close to
the SMBH by the IMBH, which spirals into the GC due to the dynamical
friction.  Our results show that, once the IMBH is formed, it brings
the massive stars to the vicinity of the central SMBH even after the
star cluster itself is disrupted.  Stars carried by the IMBH form a
disk similar to the observed disks and the core of the cluster including
the IMBH has properties similar to those of IRS13E, which is a compact
assembly of several young stars.

\end{abstract}


\section{Introduction}
Young and massive stars have been found within one parsec from the
Galactic center (GC) \citep{Krabbe95,Paumard06}. Some of them are
$\sim$1000 AU from the central SMBH. How these massive stars were brought 
to the vicinity of the SMBH has been a mystery. One possible scenario 
is the following.  A star cluster formed at a few tens of pc from the GC, 
and then spiraled in due to the dynamical friction \citep{Gerhard01}. 
Previous simulations \citep{PZ03,KM03,Kim04,GR05} have shown that the 
timescale of spiral-in of the star cluster can be short enough. 
However, how close the stars can actually approach the SMBH is not clear. 

Another possible scenario is the in-situ formation in an accretion disk
\citep{LB03,Nay07,HN08}. Giant molecular clouds fall into the GC and form 
massive gaseous disks around the central BH. Stars form in the disk 
if it becomes gravitationally unstable and results in fragmentation.
However, accretion disks have difficulty producing stars with eccentric 
orbits and 
a compact assembly of stars like IRS 13E, which is located at $\sim 0.13$ pc 
in projection from the GC and contains half a dozen young stars and probably 
an IMBH \citep{HM03,Maillard04}. \citet{Mapelli08} argued that if a gas cloud 
undergoes a very close encounter (pericenter distance of 0.01 pc) with the 
central SMBH, the tidal compression could trigger the star formation,
resulting in stars in close, bound orbits. However, how a cloud can
come that close to the GC is not clear.
On the other hand, if the star cluster has an eccentric orbit, the orbits of 
stars escaped from the cluster are also eccentric. The remnant of the core
looks like IRS 13E.

We performed a fully self-consistent $N$-body simulation in which the
internal dynamics of the cluster, that of the parent galaxy, and
interactions between cluster stars and galaxy stars are correctly
handled. In previous simulations, when the internal dynamics was
followed by an accurate $N$-body code, the parent galaxy had to be
modeled as a fixed potential with some fitting formulae for the
dynamical friction. This means that the orbital evolution is not
accurate. \citet{Fujii08} showed that
the actual orbital decay of the cluster is faster than that of previous
simulations and the main reason is stars with a mass grater
than 90\% of an initial cluster escaping from this cluster.

We describe the method of our $N$-body simulation in section 2. In
section 3 we show the results of simulations. Section 4 is for summary.

\section{Method}
\subsection{Models and Initial Conditions}
\begin{figure}
\epsscale{0.6}
\plotone{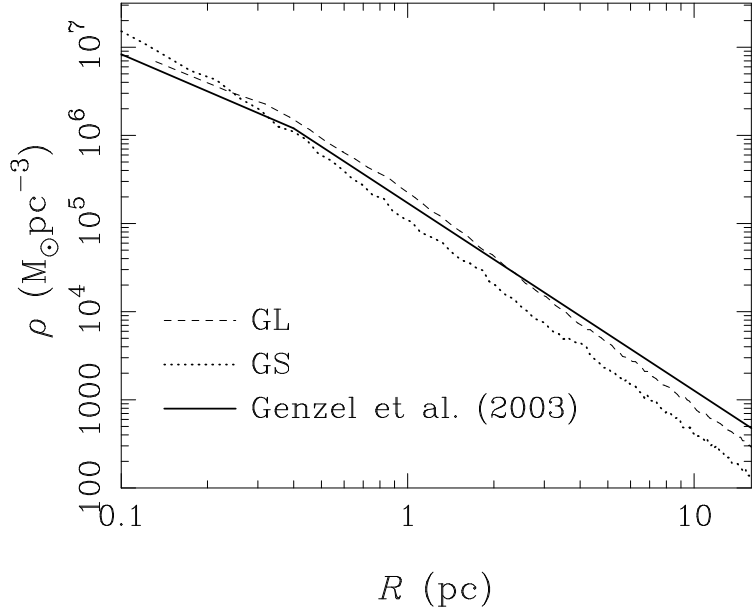}
 \caption{Density profiles for our models and the Galactic center
   \citep{Genzel03}.}
 \label{fig:density}
\end{figure}

We adopted two models for the galactic center. For both values,
we adopted a model based on a King model with the non-dimensional
central potential $W_0=10$ as galaxy models. We placed the central 
SMBH with the mass of $3.6\times 10^6 M_{\odot}$ \citep{E05} and our
galaxy models represent the central region of our Galaxy (see figure
\ref{fig:density}). The difference between two models corresponds to the
initial position of a star cluster. One (GL) is for a run 
with the initial position of the cluster 12.5 pc from the GC, and the other 
(GS) 5 pc from the GC. 
Their total masses on the real scale (excluding the SMBH) are
$5.9\times10^7 M_{\odot}$ and $2.9\times10^7 M_{\odot}$, respectively.
The number of particles is $2\times 10^6$ for both models.
GS has better mass-resolution, but a smaller half-mass radius of 9.6 pc.
GL has a larger half-mass radius of 22 pc for the farther initial
position of the cluster.
These models are summarized in table \ref{tb:modelg}.

As a model of a star cluster, we adopted a King model with non-dimensional 
central potential of $W_{0}=6$. For runs from the initial distance of
12.5 pc from the GC, we used a model with 64k particles and the total
mass of $2.1\times 10^5 M_{\odot}$ (SC64k). For runs from 5pc, we used a model 
with 32k particles and the total mass of $1.0\times 10^5 M_{\odot}$ (SC32k).
Initial mass function of stars in the clusters is a Salpeter with lower and 
upper cutoff at 1 and 100 $M_{\odot}$ \citep{Sal55}. We assigned each star a 
mass randomly chosen from the Salpeter initial mass function, irrespective 
of its position.
The tidal radii of the models are 1.1 pc for SC64k and 0.65 pc for SC32k.
They are smaller than the tidal limits at their initial positions.
The cluster models are summarized in table \ref{tb:modelsc}.

\subsection{$N$-body Simulation}
We used the Bridge code \citep{Fujii07} to handle the interaction
between the parent galaxy and the star cluster fully self-consistently. 
The Bridge scheme is a tree-direct hybrid scheme. 
Only the internal motion of the star cluster is calculated by the direct 
scheme with high accuracy, and all other interactions are calculated by 
the tree algorithm. The splitting between the direct part and tree part 
is through the splitting of the Hamiltonian in a way similar to the 
mixed variable symplectic \citep{KYN91,WH91}. With the Bridge scheme, we 
can treat a large-$N$ system with embedded small-scale systems fully 
self-consistently and accurately.

We used two sets of numerical parameters for each galaxy model.
They are summarized in table \ref{tb:param}.
In our model, the softening length between galaxy particles and the SMBH is 
0.2 pc. Therefore, our simulation has the resolution limit around 0.2 pc for 
the motion of the star cluster and cluster stars within the parent galaxy.
We used the opening angle $\theta = 0.75$ with the center-of-mass 
approximation for the tree. 
The simulation is performed using GRAPE6 \citep{Makino03}.

\subsection{Stellar Collisions and Evolutions}
In our simulation, we adopted collisions of stars in a star cluster and
formation of an IMBH in the cluster. Recent simulations showed that in a 
dense star cluster, runaway collisions of stars form a very massive star 
\citep{PZ99,PZM02,PZ04,FGR06}. If it is massive enough, it will collapse 
into an IMBH \citep{FWH01,Heger03}. While the massive star grows through 
collisions, it loses its mass due to the stellar wind. Very massive stars
lose their mass rapidly \citep{Belkus07}. However, these processes are not 
well understood for stars more massive than $1000M_{\odot}$. Therefore, we 
treated the collision criterion and the mass-loss rate as parameters. 
In this section, we describe our model of stellar collision and evolution
and discuss the difference caused by these parameters.

If the core collapse occurs in the star cluster, collisions between
stars in the core occur rather frequently. We let two stars merge in the 
star cluster if they approach to twice the sum of stellar radii, which 
is also adopted in \citet{PZ99}. 
If two stars approach, the tidal capture occurs: the tidal force exerts 
stellar oscillations and the stars lose their orbital energy and become binary 
\citep{FPR75,PT77,McMillan87}. The critical distance for the tidal capture 
is around 3-4 times the radius of the star for two identical stars.
When a tidally captured binary is actually formed, it would merge fairly
quickly because of perturbations from nearby stars.
For the radii of stars, we adopted the result of \citet{HPT00}
(here after HPT00). 
The structure of very massive stars with $100<M<1000M_{\odot}$ is 
investigated by \citet{Ishii99} and \citet{Yungelson08}, while that of stars
more massive than 
$1000M_{\odot}$ is not. We used the results of HPT00 for very massive stars 
with $>100M_{\odot}$. Massive stars ($>100M_{\odot}$) have a core-halo structure
\citep{Ishii99}. The result of HPT00 we adopted is close to the size of
the core in \citet{Ishii99}.
Therefore, we adopted twice the sum of the radii as the collision criterion.
To see the effect of the different criterion, we also performed simulations 
using the sum of the stellar radii as the collision criterion
\citep{PZM02,PZ04,FGR06}.

We also take into account the mass loss due to the stellar wind for very 
massive stars. 
\citet{Belkus07} have investigated the mass-loss rate for massive stars 
with $<1000 M_{\odot}$. However, the stellar evolution of stars with 
$>1000M_{\odot}$ is not well understood. Furthermore, merged stars show a 
evolution different from that of single stars \citep{Suzuki07}. Therefore, 
we treat the mass-loss rate as a parameter.

\begin{figure}
\epsscale{0.5}
\plotone{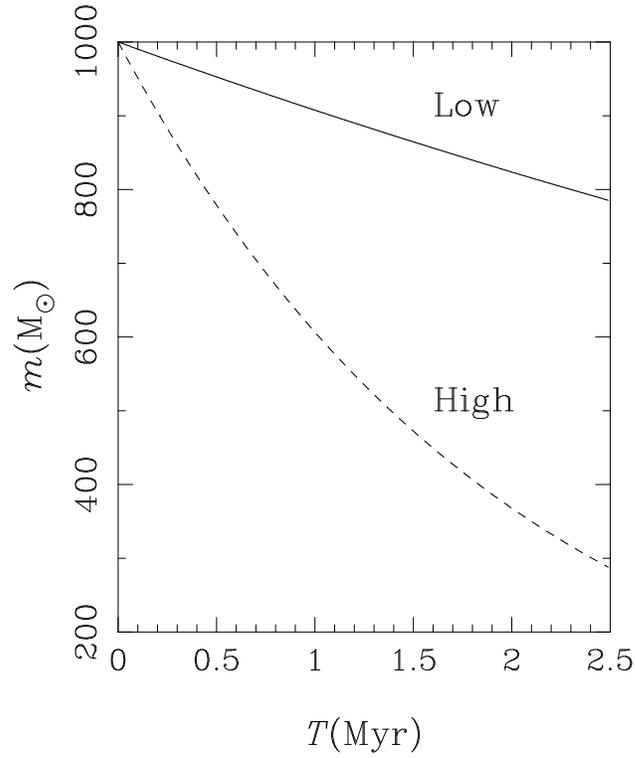}
 \caption{Evolution of the mass for stars with initial mass 
 $=1000M_{\odot}$. The solid and dashed curves are the results with low 
 mass-loss rate (see equation (\ref{eq:mlrate}) in the text) and high 
 mass-loss rate (five times as that of equation (\ref{eq:mlrate})), 
 respectively.}
 \label{fig:mloss}
\end{figure}

Our model for the stellar wind is based on line-driven
winds developed by Castor et al. (1975; hereafter CAK).
We used the formalism of \citet{Owocki04}.  The
mass-loss rate of the CAK model is given by
\begin{eqnarray}
\dot{M}_{*} = \frac{L_{*}}{c^2}\frac{\alpha}{1-\alpha}
\left[ \frac{\bar{Q}\Gamma_{e}}{1-\Gamma_{e}} \right]
^{(1-\alpha)/\alpha},
\end{eqnarray}
where $L_{*}$ is the luminosity of the star, $\Gamma_e$ is the Eddington
parameter, and $\alpha$ and $\bar{Q}$ are the power index and
normalization of the line opacity distribution.
The Eddington parameter is given by
\begin{eqnarray}
\Gamma_{e} = \frac{\kappa_{e}L_{*}}{4\pi G M_{*}c},
\end{eqnarray}
where $M_{*}$ is the mass of the star, $G$ is the gravitational
constant, $c$ is the speed of light, and $\kappa_e$ is the opacity.
For fully ionized plasmas with hydrogen mass fraction $X$, the opacity
is given by $\kappa = 0.2(1+X)$ cm$^{2}$ g$^{-1}$, where we assumed 
$X=0.7$. For very massive
stars, the luminosity is proportional to the mass because in such
massive stars, the contribution of radiation to the total pressure is
very large \citep{Marigo03}.  Hence, we assumed $\dot{M_{*}} \propto M_{*}$,
here $M_{*}$ is the mass of the star.
We adopted $\alpha = 0.5$(CAK), $\bar{Q}=10^3$ \citep{Gayley95},
$L_{*} = 3.2\times 10^4 (M_{*}/M_{\odot})(L_{\odot})$ (Suzuki, private
communication), which is close to the Eddington luminosity.
From these values, we obtained
\begin{eqnarray}
\dot{M}_{*}=9.66\times 10^{-8} \left(\frac{M_{*}}{M_{\odot}}\right)
 (M_{\odot}/{\rm yr}).
\label{eq:mlrate}
\end{eqnarray}
On the other hand, the recent results \citep{Belkus07} show a higher 
mass-loss rate. 
Therefore, we also adopted a mass-loss rate five times higher,
similar to the result of \citet{Belkus07}. The evolution of the
mass for stars with 1000 $M_{\sun}$ initial mass is shown in figure
\ref{fig:mloss}.
We neglected the mass loss of the stars less than 300 $M_{\odot}$ because
they are small enough. 

We assumed that very massive stars formed by runaway collisions form
black holes (BHs) at the end of their main-sequence lifetime 
\citep{FWH01,Heger03}. For the lifetime, we adopted the results of
HPT00 for stars with masses less than 100 $M_{\odot}$,  
interpolated the results of \citet{Belkus07} for stars with masses up to 
1000 $M_{\odot}$, and extrapolated them for more than 1000 $M_{\odot}$.
We assigned new ages to stars born by stellar collisions using the following 
formalism of \citet{MvdH89}:
\begin{eqnarray}
t_{\rm age}(m_{1}+m_{2}) = \frac{m_{1}}{m_{1}+m_{2}} 
\frac{\tau _{\rm ms}(m_{1}+m_{2})}{\tau _{\rm ms}(m_{1})} t_{\rm age}(m_{1}),
\label{eq:lifetime}
\end{eqnarray}
where $m_{1}$ and $m_{2}$ are the masses of stars ($m_{1} > m_{2}$),
$t_{\rm age}$ is the age of the star, and $\tau _{\rm ms}$ is the
main-sequence lifetime.
The new star looks rejuvenated.
However, the assumption used to derive equation (\ref{eq:lifetime}) is that 
the convection core of the primary  will occupy the core of the merger
with the mass and composition unchanged. \citet{Suzuki07} and also
\citet{Gaburov08}
found that in the case of unequal-mass merger the core of the
secondary will first occupy the center of the merger, resulting in
significant mixing-in of hydrogen into the new convection core. So
equation (4) might underestimate the lifetime of the merger.
We assumed that the massive star directly collapses to an BH.
The final evolution of the very massive star depends on its helium core mass 
at the end of its life. If it is more massive than $\sim130M_{\odot}$, it 
collapses directly to an BH without mass loss \citep{FWH01}.

\begin{figure}
\epsscale{0.6}
\plotone{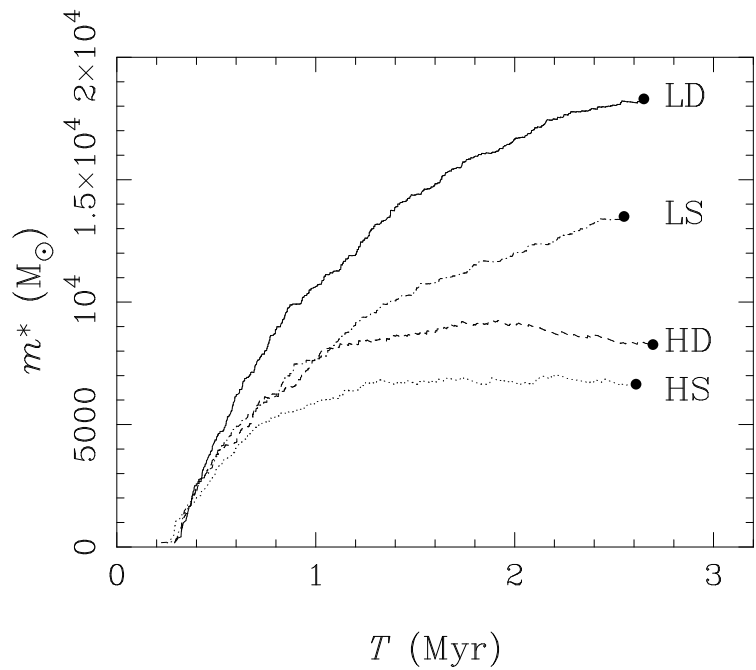}
 \caption{Evolution of the most massive star in the star cluster. 
   The model of star cluster is SC64k. Black points show the time when 
   IMBHs formed. "L" and "H" mean low- and high-mass loss rates, and "D" 
and "S" denote the collision radius. See table \ref{tb:run1} for details.}
 \label{fig:m_star}
\end{figure}

We performed simulation for isolated star clusters and investigated the 
difference caused by the parameters. The model of star cluster is SC64k.
The runs are summarized in table \ref{tb:run1}. Each parameter has two values.
The results are shown in figure \ref{fig:m_star}.
The difference due to $r_{\rm coll}$ is smaller than that due to the mass-loss 
rate because the stellar radii is smaller than the radius, but it becomes 
larger when the mass grows. The mass-loss rate significantly affects the
final mass. In any case, however, the mass accretion rate due to stellar 
collisions was higher or almost the same as the mass-loss rate. The mass loss
will not prevent the formation of an IMBH in the star cluster.

\subsection{Runs}
We performed fully self-consistent $N$-body simulations of a star cluster 
and its parent galaxy system. Runs are summarized in table \ref{tb:run2}.
For the mass-loss rate and collision criterion, we adopted two extreme
cases. They are upper and lower limits of the mass of IMBHs.

\section{Results}
\subsection{Evolution of Star Cluster and IMBH}
\begin{figure}
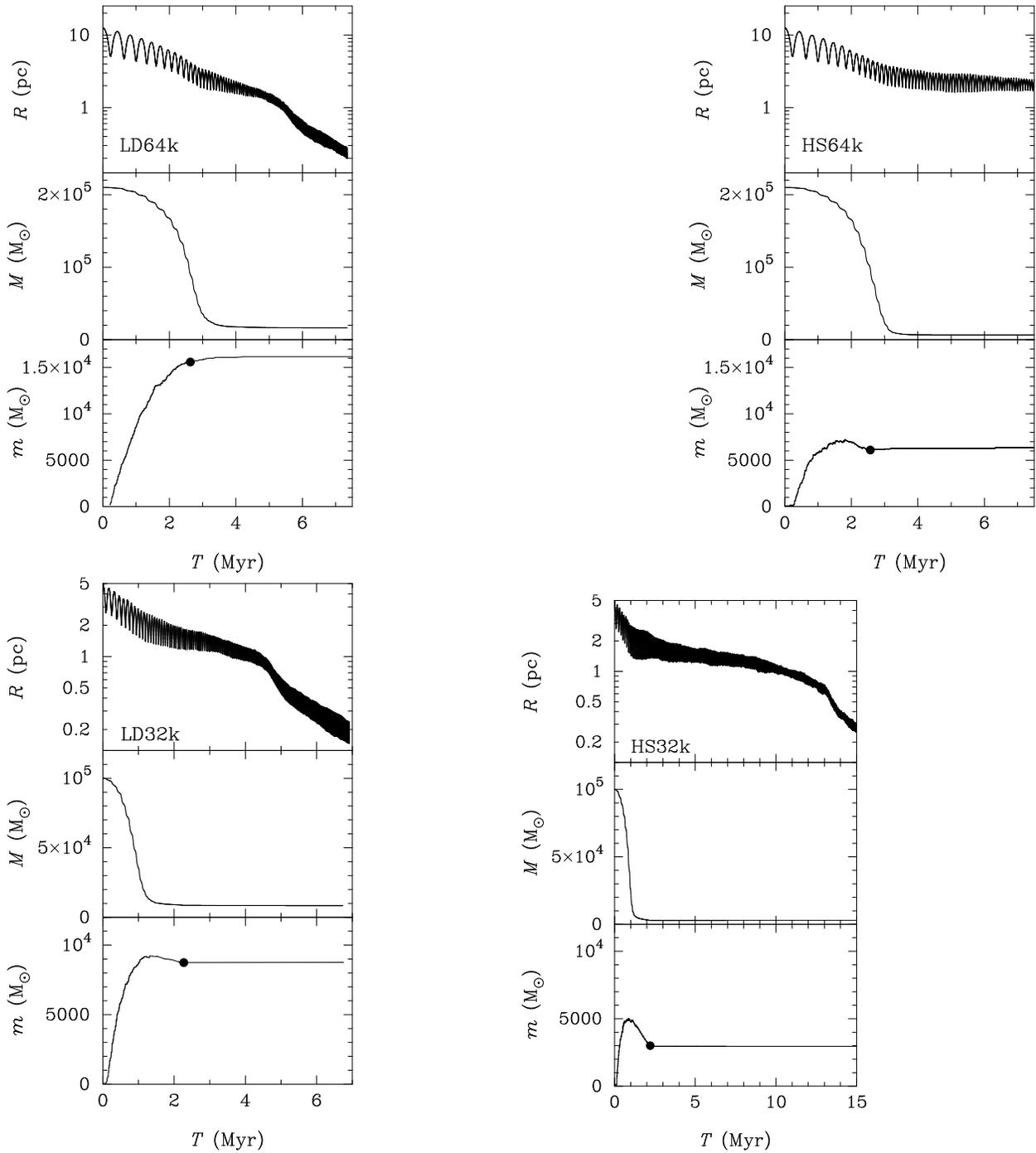

\epsscale{0.75}
\plottwo{f4a.eps}{f4b.eps}
\plottwo{f4c.eps}{f4d.eps}
 \caption{Evolution of star clusters. Top and middle panels show the
 orbital and bound-mass evolution of star clusters, respectively. Bottom
 panel shows the evolution of the most massive star in the star
 clusters. The black dot shows the time when the massive star became an IMBH.}
 \label{fig:star_cluster}
\end{figure}

The orbit of the star cluster decayed due to the dynamical friction and
the cluster was disrupted by the tidal force.  
The evolutions of star clusters are shown in figure \ref{fig:star_cluster}.
Top and middle panels show the orbital and bound-mass evolution of the star 
clusters. Bottom panels show the mass of the most massive star in the 
star cluster. Black points in the bottom panels show the times when IMBHs
formed.

In each case, when the pericenter distance becomes less than 2 pc,
the star clusters are almost completely disrupted. 
The disruption time, $r_{\rm dis}$, is summarized in table \ref{tb:results}. 
We defined the disruption as the time when the number of stars bound to the
cluster becomes less than ten.

Before the cluster was completely disrupted, an IMBH formed in the
cluster through the runaway collisions of stars.  A massive star grew
via repeated collisions and finally the mass of the massive star
reached 3000-16000 $M_{\odot}$. The evolution of the massive star is shown
in bottom panels of figure \ref{fig:star_cluster}. The IMBH mass strongly 
depends on the mass-loss rate. The time when the IMBH formed was earlier 
than the disruption time except for the run HS32k.
These results are summarized in table \ref{tb:run2}.

After the disruption of the star cluster, the IMBH formed
sinks into the GC on the time scale of the Chandrasekhar dynamical friction
\citep{Ch43}. The time scale of the orbital evolution due to the dynamical
friction is proportional to the mass of the star cluster (IMBH). The IMBH 
mass of LD64k is around three times more massive than that of HS64k. The 
orbital 
evolution of LD64k is roughly three times faster than that of HS64k. 
On the other hand, the orbital decay of the cluster looks faster than the 
time scale. This is because the dynamical friction was enhanced due to
the stars escaping from the cluster \citep{Fujii06}.

The orbital evolution of the star cluster became slower and slower at the 
end of the simulation. This is because we adopted a large softening 
length between the SMBH and galaxy particle, which is 0.2 pc. If the
softening length were smaller, the cluster would approach closer to the
GC and carry resonant stars. We will 
perform a simulation with a smaller softening length and report the results 
in a forthcoming paper.

\begin{figure}[htbp]
\epsscale{1.0}
\plotone{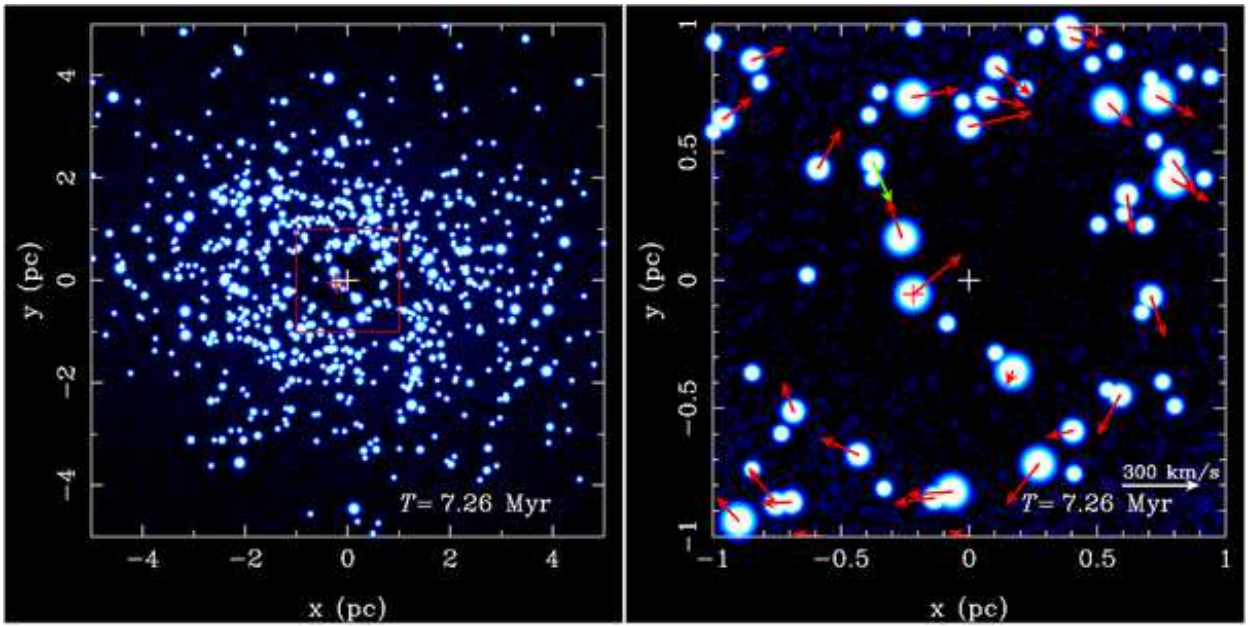} 
 \caption{
   Projected distribution of stars at $T=7.26$ Myr for LD64k. 
 The orbital plane of the star cluster has an  inclination of
 $i=127^{\circ}$ with respect to the plane of the sky, with a half-line
 of ascending  nodes at $\Omega = 99^{\circ}$ east of north. These
 values are used to mimic the result of
 \citet{Paumard06}.  Right panel shows the central region
 shown by a red square in left panel.  Arrows in right panel show the proper
 motion of stars.  Red one shows clockwise orbit and green shows
 counter-clockwise.  Red and white crosses show the positions of the
 IMBH and SMBH, respectively.  
 }
 \label{fig:snapshot}
\end{figure}

Figure \ref{fig:snapshot} shows the projected distribution of stars 
at the end of the run (7.25Myrs) of LD64k.  Only
stars which were originally members of the star cluster are plotted.
The arrows in the right panel show the proper motions of the
stars.  Red ones move clockwise and green ones move counterclockwise.
This figure looks very similar to K-band images of the GC
\citep[cf.][figure 2]{Lu06}.  In particular, they reached a 
distance of $\sim 0.2$ pc from the GC. We can see that, even though
the cluster is almost completely disrupted at 2pc, there  are a number
of stars which were brought much closer to the central SMBH.

\subsection{Trojan Stars}
\begin{figure}
\epsscale{0.8}
\plotone{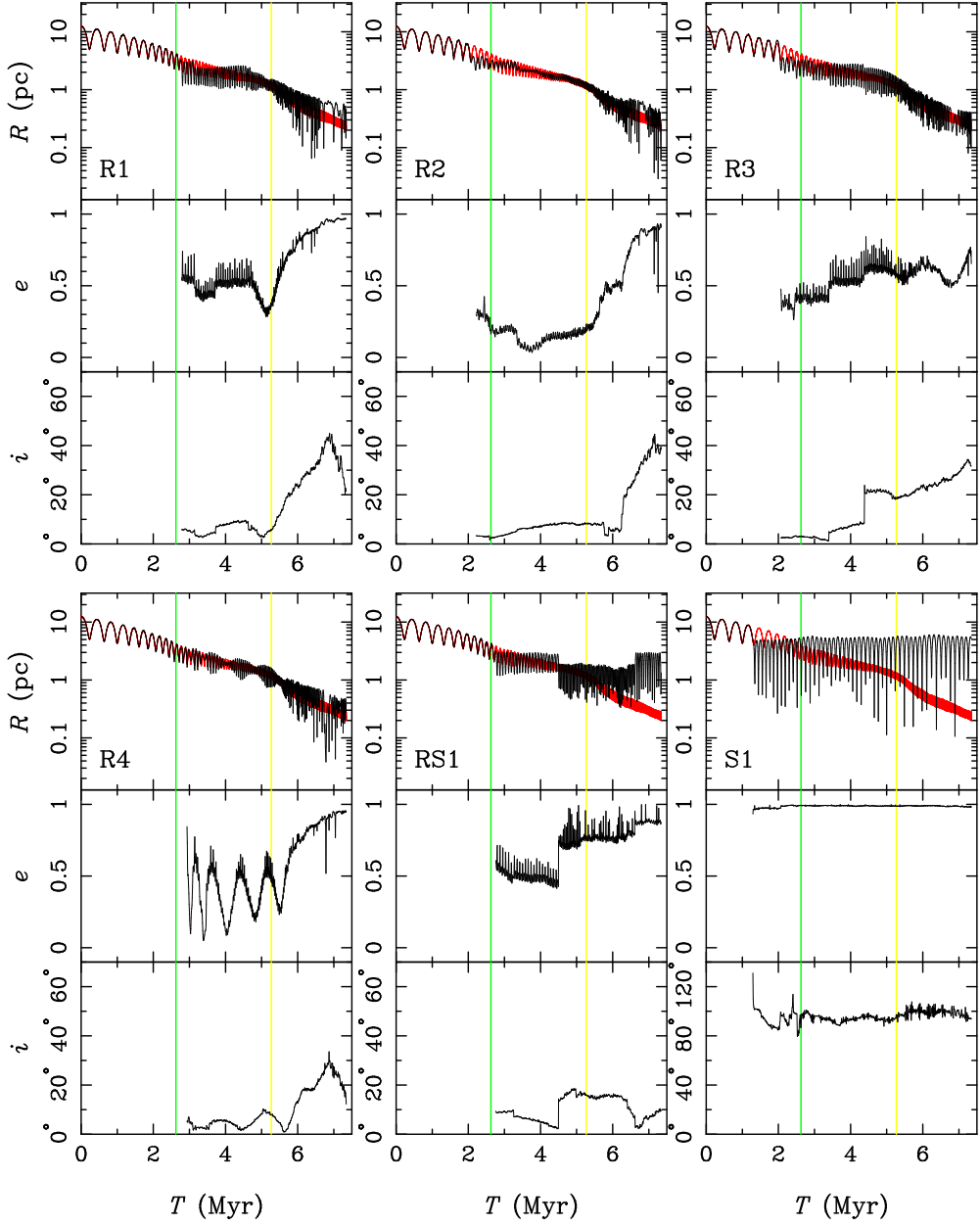}
 \caption{Evolution of position, eccentricities, and inclinations of stars
 escaped from the star cluster for LD64k.  Red curve shows the orbital evolution
 of the star cluster or the IMBH formed in the cluster.  Green and
 yellow lines show the time when the IMBH
 was formed and when the star cluster was disrupted (less than ten
 stars are bound to the cluster), respectively. }
 \label{fig:evolution}
\end{figure}

We found that many stars were carried near the GC though almost all
stars became unbound at around 1-2 pc from the GC.
Figure \ref{fig:evolution} shows how these stars were brought near the
central SMBH.  Red solid curves show the orbit of the star cluster.
Here, we plotted the trajectory of six stars of LD64k. 
All of them have escaped
from the cluster by 3Myrs, as can be seen from the separation of black
and red curves. One would imagine that their orbits do not evolve once
they have escaped from the cluster.  However, actually these stars
``follow'' the spiral-in of the cluster remnant. This behavior occurs
due to the 1:1 resonance with the IMBH formed in the cluster. These
stars have a pericenter distance of less than 0.1pc at the end of the
simulation. 

\begin{figure}
\epsscale{0.8}
\plotone{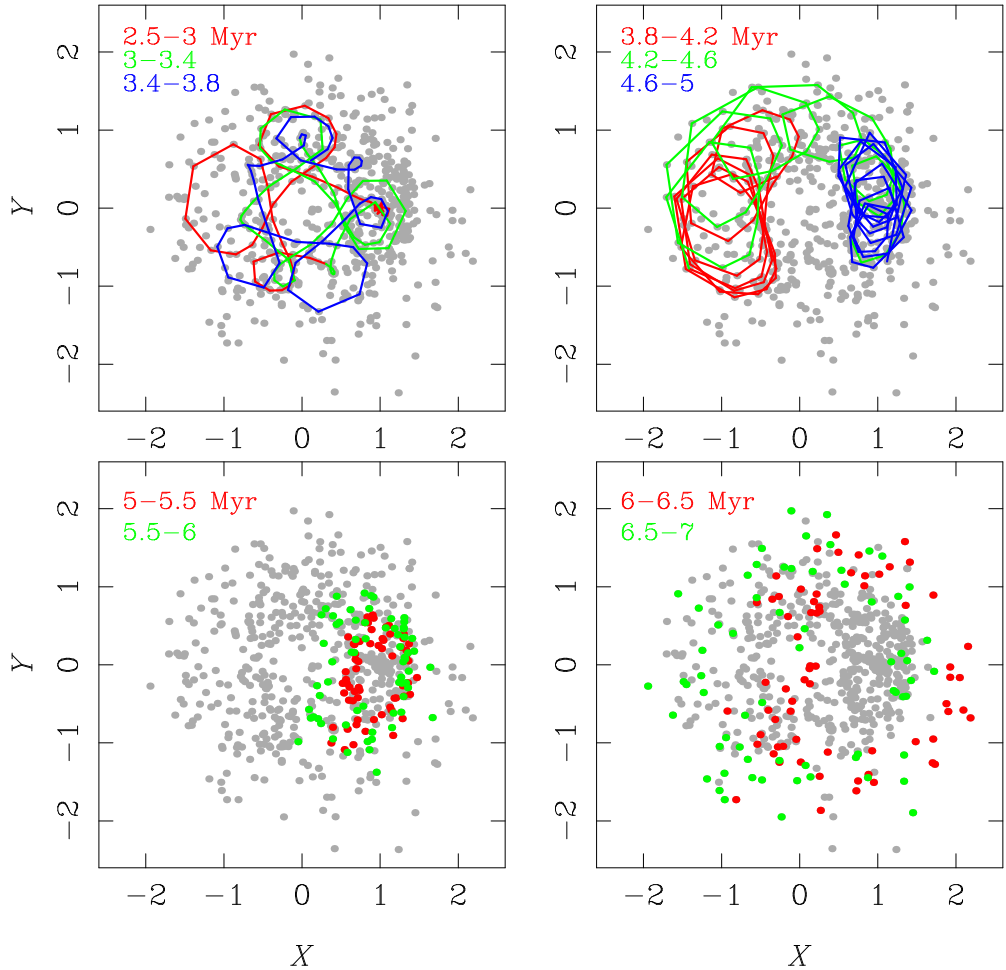}
 \caption{Orbit of star R1 in a rotational frame, where the IMBH is
 fixed at (1.0, 0.0). The SMBH lies on the origin. The distance of the
 star from the SMBH is normalized by the distance between the SMBH and
 the IMBH. Gray dots show all positions of the star obtained from snapshots.
 Colored curves or dots show the orbit or position of the star within
 time spans given in each panel.}
 \label{fig:orbit}
\end{figure}

In figure \ref{fig:evolution}, stars R1-R4 were brought to the GC by
the 1:1 mean motion resonance.  Figure \ref{fig:orbit} shows the orbit
of R1 in a rotating frame where the distance between the SMBH and IMBH
is scaled to unity.  Star R1 escaped from the cluster at $T\simeq 2.5$ Myr and
was orbiting around the SMBH until $T\simeq 3.8$ Myr. At $T\simeq 3.8$
Myr, it was captured into the resonance.  Its orbit was a horseshoe
orbit for $T\simeq 3.8$ Myr to $T\simeq 5$ Myr.  From $T\simeq 5$
Myr to $T\simeq 6.5$ Myr, the orbit of R1 is that of a retrograde
quasi-satellite.  After that, it escaped from the resonance. Stars R2
through R4 show similar behaviors.  Their distances from the GC have
become smaller by a factor of five or more while they are in the
resonance with the IMBH.  Some of them escape from the resonance on
the way to the GC.

Some stars are scattered by the IMBH during the resonance and change
their orbit into those approaching the GC.  We categorized such
stars into a Resonance and Scatter group (RS), and show one example in
figure \ref{fig:evolution}.  In this case, its semi-major axis does 
not decrease gradually, but suddenly through close encounters with the
IMBH (see figure \ref{fig:evolution}, middle-bottom panel).  These
resonant stars (R and RS) make up more than 90 \% of stars whose
pericenter distances, $r_{\rm p}$, are less than 0.5 pc from the GC
(Table \ref{tab:origin}).

The right-bottom panel in figure \ref{fig:evolution} shows the orbital
evolution of a star which escaped from the star cluster by a
slingshot occurring in the star cluster. We categorized such stars into 
a Slingshot group (S).  Some of them come very close to the GC.  
Moreover, these stars have very high inclinations 
(some have retrograde orbits) to the orbital plane of the
star cluster because the escape direction from the star cluster is isotropic.
They may be the origin of the stars which have high inclinations in the GC.
Table  \ref{tab:origin} summarizes these results.  We can see that
the 1:1 resonance with the IMBH is the main mechanism which
brings the stars close to the GC.

\begin{figure}
\epsscale{0.8}
\plotone{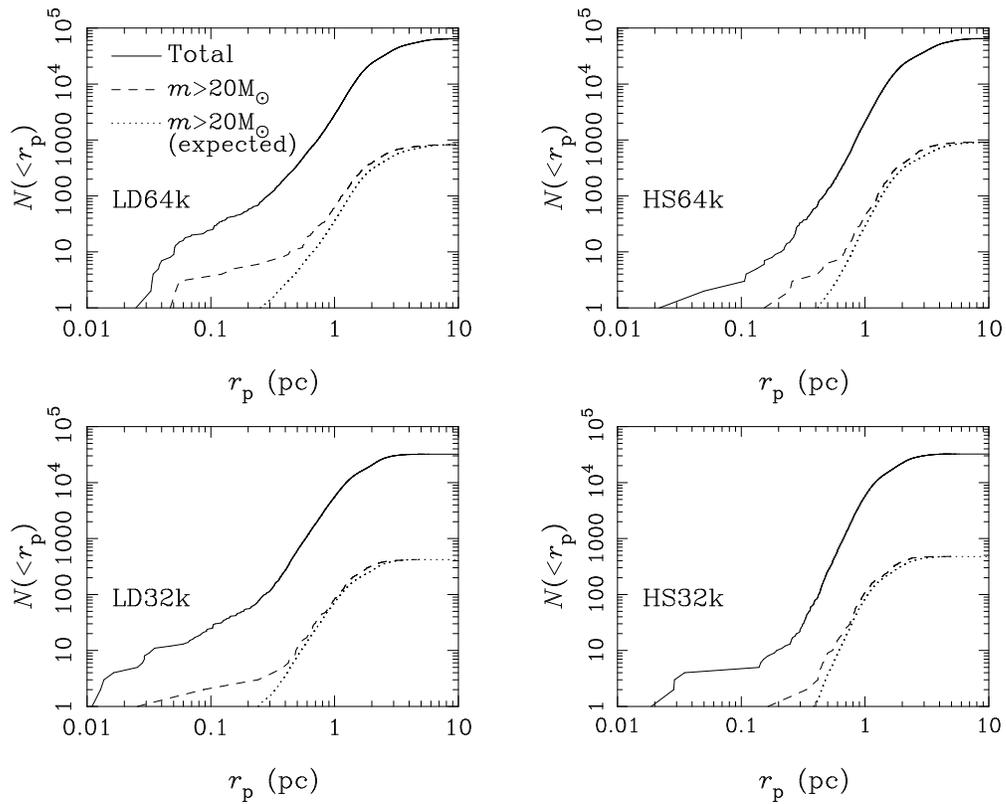}
 \caption{Cumulative number of escaped stars as a function of peri-center 
   distance at the end of simulations. Solid curves include all stars and 
   dashed curves only massive stars ($>20M_{\odot}$). Dotted curves show the 
   expected number of stars if the fraction of stars is the same, 
   irrespective of $r_{\rm p}$.}
 \label{fig:n_rp}
\end{figure}

Figure \ref{fig:n_rp} shows the cumulative number of escaped stars as a 
function of pericenter distance at the end of simulations. 
Solid curves include all stars and dashed curves only massive stars 
($>20M_{\odot}$). Dotted curves show the expected number of stars if the 
fraction of stars is independent of $r_{\rm p}$. 
We can see that there is a much larger number of massive stars in the central 
sub-parsec region than expected from the initial mass function.

In the case of run LD64k,
there are 16 massive stars ($m>20M_{\odot}$) with the pericenter
distance less than 0.5 pc.  Five of the 16 stars reached within 0.2 pc
and three reached within 0.1 pc (see table \ref{tab:m_stars}). More than
10 \% of stars with $r_{\rm p}<0.5$ pc have masses $20 M_{\sun}$. If 
the distribution of stars does not depend on their masses, the expected 
number of massive stars for the radius of 0.1pc is around 0.3. 
If there is no mass-dependent evolution effect, the chance that
we find more than three stars with $r_{\rm p} < 0.1$ pc is $10^{-3}$.
In the other cases, a few persent of stars near the GC was massive at 
$r_{\rm p}=0.5$ (pc) and the fraction is higher at more deeper region.
In the initial mass function, only 0.5\% stars have mass more than 
$20M_{\odot}$ and $\sim0.1$\% after runaway collisions because many massive 
stars are used up for the formation of the IMBH. 

Figure \ref{fig:n_rp} also shows that the number of the Trojan stars 
depends both on the mass of IMBH and the spiral-in timescale of IMBH.
The number of resonant stars decreases when the IMBH mass is smaller.
However, as the IMBH spirals in towards the GC in 5-6 Myrs, the number of
Trojan stars is not much different from that in our standard run (LD64k).

These results imply that star clusters selectively 
carry massive stars close to the GC. Massive stars sink to the center 
of their parent star clusters due to the mass segregation, while the 
stripping by the tidal force removes less massive stars from the outskirts 
of star clusters.  As a result, massive stars tend to remain in star 
clusters and are carried to a few parsec from the GC before they escape 
from the cluster. Slingshot stars also include many massive stars because 
slingshots occur at the center of the star cluster where massive stars are 
gathering. Thus, star cluster scenarios can naturally explain
why massive stars lie within the central parsec.

\subsection{The Remnant of the Core and IRS 13E}
\begin{figure}
\epsscale{0.6}
\plotone{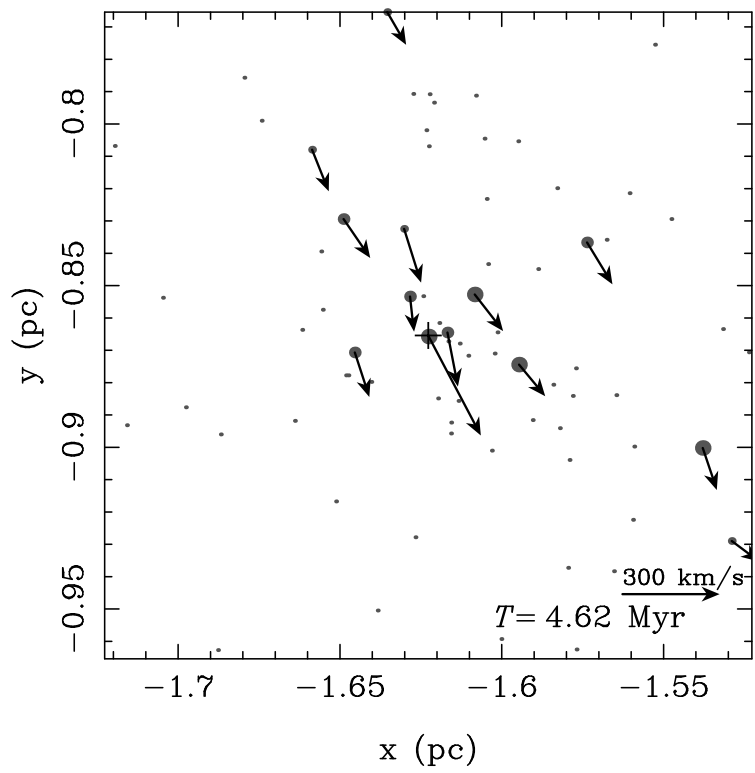}
 \caption{Snapshot of the remnant of the core. Vectors show the proper 
   motion of stars for massive stars ($>20M_{\odot}$). It is similar to 
   IRS 13E \citep[see figure 2 in][]{Maillard04}.}
 \label{fig:core}
\end{figure}

At the end of the simulation, only one massive star is bound to the IMBH.
However, several stars were bound till $T=6$ Myrs for LD64k. These bound 
stars look like IRS 13E, which is located at $\sim 0.13$ pc in projection 
from the GC and contains half a dozen massive and young stars within 
$\sim 0.01$ pc. Because of their very similar proper motions, it has been 
suggested that they are bound to an IMBH and IRS 13E is the remnant of 
a star cluster containing an IMBH with $\sim 10^4 - 10^5 M_{\odot}$ 
\citep{HM03,Maillard04}. 
Figure \ref{fig:core} shows the proper motion of stars bound to the IMBH 
at $T=4.62$ Myr for LD64k. The tidal radius is 0.15 pc and larger than the 
frame, but a few stars are within 0.01 pc. This is similar to the observation
of the IRS 13E \citep[see figure 2 in][]{Maillard04}.

\section{Summary}
Using $N$-body simulation, we showed that many young and massive stars
are carried to the GC by a star cluster due to the 1:1 mean motion
resonance with an IMBH which is formed in the cluster.  In addition,
we found that slingshots in the star cluster throw stars into 
orbits which pass near the GC.  These orbits have very high
inclinations and are sometimes retrograde orbits.  They are new channels
which carry young stars to the central parsec. Our simulation demonstrated
the existence of massive stars and we explained why they form a disk-like
structure.  The possible existence of two counter-rotating disks might
suggest that two clusters have spiraled in.  The interaction and
resonance between stars and IMBHs from multiple clusters might be
responsible for the existence of stars which are very close to the
central SMBH.

\acknowledgments

The authors thank Takeru Suzuki for the data of very massive stars,
Eiichiro Kokubo and Yusuke Tsukamoto for helpful discussions, and 
the referee, Stefan Harfst, for useful comments on the manuscript.
M. F. is financially supported by Research Fellowships of JSPS for Young
Scientists.  
This research is partially supported by
the Special Coordination Fund for Promoting Science and Technology
(GRAPE-DR project), Ministry of Education, Culture, Sports, Science and
Technology, Japan.
Part of calculations were done using the
GRAPE system at the Center for Computational Astrophysics (CfCA) of
the National Astronomical Observatory of Japan.

\onecolumn

\begin{table}[htbp]
\begin{center}
\caption{Models for the galaxy\label{tb:modelg}}
\begin{tabular}{ccccccccccc}
\tableline
\tableline
& King $W_0$& $N$ & $M_{\rm G}({\rm M_{\odot}})$ & $m ({\rm M_{\odot}})$ & $M_{\rm BH}({\rm M_{\odot}})$ &
$r_{\rm h}$ (pc) \\
\tableline
GL & 10 & $2 \times 10^6 $ & $5.8 \times 10^7$ & 29  & $3.6 \times 10^6$ &  
22  \\
GS & 10 & $2 \times 10^6 $ & $2.9 \times 10^7$ & 15 & $3.6 \times 10^6$ &
 9.6  \\
\tableline
\end{tabular}
\end{center}
\end{table}

\begin{table}[htbp]
\begin{center}
\caption{Models for the star cluster\label{tb:modelsc}}
\begin{tabular}{cccccccccc}
\tableline
\tableline
& King $W_0$& $N$ & $M_{\rm SC}({\rm M_{\odot}})$ & $r_{\rm c}$ (pc) &
$r_{\rm h}$ (pc) & $r_{\rm t}$ (pc)\\
\tableline
SC64k & 6 & 65536 & $2.0 \times 10^5$ & $5.9 \times 10^{-2}$
 & $1.6 \times 10^{-1}$ & $1.1$\\
SC32k & 6 & 32768 & $1.0 \times 10^5$ &  $3.5 \times 10^{-2}$
 & $9.6 \times 10^{-2}$ & $6.5  \times 10^{-1}$\\
\tableline
\end{tabular}
\end{center}
\end{table}

\begin{table}[htbp]
\begin{center}
\caption{Parameters\label{tb:param}}
\begin{tabular}{cccccc}
\tableline
\tableline
 & $\epsilon _{\rm SC}$ (pc) & $\epsilon _{\rm G}$ (pc) & $\epsilon _{\rm SC-BH}$ (pc)& $\epsilon _{\rm G-BH}$ (pc) & $\Delta t_{\rm tree}$ yr\\
\tableline
P1 & 0.0 & $4.9\times 10^{-2}$ & $2.5\times 10^{-2}$ & $2.5\times 10^{-1}$ & $2.4\times 10^2$\\
P2 & 0.0 & $2.3\times 10^{-2}$ & $2.4\times 10^{-2}$ & $2.4\times 10^{-1}$ & $1.1\times 10^2$ \\
\tableline
\end{tabular}
\end{center}
\end{table}

\begin{table}[htbp]
\begin{center}
\caption{Runs for isolated star clusters\label{tb:run1}}
\begin{tabular}{ccccccc}
\tableline
\tableline
  & $\dot{m}$ & $r_{\rm coll}$ & $M_{\rm IMBH} (M_{\odot})$ & $M_{\rm lost} (M_{\odot})$\\
\tableline
LD & Low\tablenotemark{1} & Double\tablenotemark{3} & $1.8\times 10^4$ & $2.9\times 10^3$\\
LS & Low & Single\tablenotemark{4} & $1.4\times 10^4$ & $2.0\times 10^3$\\
HD & High\tablenotemark{2} & Double & $8.3\times 10^3$ & $9.0\times 10^3$\\
HS & High & Single & $6.7\times 10^3$ & $6.8\times 10^3$\\
\tableline
\end{tabular}
\tablenotetext{1}{$\dot{m}=9.7\times10^{-8}m$ ($M_{\odot}$/yr)}
\tablenotetext{2}{$\dot{m}=5.0\times10^{-7}m$ ($M_{\odot}$/yr)}
\tablenotetext{3}{$r_{\rm coll} = r_{1} + r_{2}$}
\tablenotetext{4}{$r_{\rm coll} = 2(r_{1}+r_{2})$}
\end{center}
\end{table}

\begin{table}[htbp]
\begin{center}
\caption{Runs\label{tb:run2}}
\begin{tabular}{ccccccc}
\tableline
\tableline
 & Galaxy & Star cluster & $\dot{m}$ & $r_{\rm coll}$ & Parameter & $R_{0}$ (pc) \\
\tableline
LD64k & GL & SC64k & Low\tablenotemark{1} & Double\tablenotemark{3} & P1 & 12.5 \\
HS64k & GL & SC64k & High\tablenotemark{2} & Single\tablenotemark{4} & P1 & 12.5\\
LD32k & GS & SC32k & Low & Double & P2 & 5 \\
HS32k & GS & SC32k & High & Single & P2 & 5 \\
\tableline
\end{tabular}
\tablenotetext{1}{$\dot{m}=9.7\times10^{-8}m$ ($M_{\odot}$/yr)}
\tablenotetext{2}{$\dot{m}=5.0\times10^{-7}m$ ($M_{\odot}$/yr)}
\tablenotetext{3}{$r_{\rm coll} = r_{1} + r_{2}$}
\tablenotetext{4}{$r_{\rm coll} = 2(r_{1}+r_{2})$}
\end{center}
\end{table}

\begin{table}[htbp]
\begin{center}
\caption{Results\label{tb:results}}
\begin{tabular}{ccccc}
\tableline
\tableline
 & $t_{\rm IMBH}$(Myr) & $M_{\rm IMBH} (M_{\odot})$ & $M_{\rm lost} (M_{\odot})$ & $t_{\rm dis}$\tablenotemark{1} (Myr)\\
\tableline
LD64k & 2.6 & $1.6\times 10^4$ & $2.4\times 10^3$ & 5.3 \\
HS64k & 2.6 & $6.3\times 10^3$ & $6.5\times 10^3$ & 4.3 \\
LD32k & 2.3 & $8.8\times 10^3$ & $1.5\times 10^3$ & 2.7 \\
HS32k & 2.2 & $3.0\times 10^3$ & $4.1\times 10^3$ & 1.7 \\
\tableline
\end{tabular}
\tablenotetext{1}{Disruption time of star clusters, when the number of bound stars becomes less than ten.}
\end{center}
\end{table}

\begin{table}[htbp]
\begin{center}
\caption{The number of stars within 0.5 pc for LD64k.\label{tab:origin}}
\begin{tabular}{c|cccc|c}
\hline \hline
Pericenter distance  & Bound & Slingshot &
 Resonance & Resonance \& Scatter & total\\
\hline
$r_{\rm p}<0.5$ pc & 2 & 27 & 280 & 45 & 354\\
$r_{\rm p}<0.2$ pc & 0 & 4 & 52 & 0 & 56\\
$r_{\rm p}<0.1$ pc & 0 & 0 & 23 & 0 & 23\\
\hline
\end{tabular}
\end{center}
\end{table}

\begin{table}[htbp]
\begin{center}
\caption{The number of massive stars within 0.5 pc for LD64k.\label{tab:m_stars}}
\begin{tabular}{c|cccc|c}
\hline \hline
Pericenter distance  & Bound & Slingshot &
 Resonance & Resonance \& Scatter & Total\\
\hline
$r_{\rm p}<0.5$ pc & 1 & 5 & 7 & 3 & 16\\
$r_{\rm p}<0.2$ pc & 0 & 1 & 4 & 0 & 5\\
$r_{\rm p}<0.1$ pc & 0 & 0 & 3 & 0 & 3\\
\hline
\end{tabular}
\end{center}
\end{table}

\end{document}